# Cooperative game in a two –lane traffic flow

Chi-Yu Wang,[a] Li-Hu Wang,[b] Ruo-Hang Chen[c]
Guangxi Normal University, College of Physical Science and Technology, No 15, Yu-Cai Road, Guilin City, Guangxi, PRC, 541004

Abstract: We use cellular automata model to study the cooperation between cyclists. In the two-lane model, cyclists can change lanes. Even there is someone on the back they will take a cooperative attitude. It means that they will be in a same lattice. Simulation carried out under the open borders. Simulation results show that the density in a certain area appeared surge. When the $\alpha$ (or $\beta$) is constant and $\beta$ (or $\alpha$) changes in isometric, the distance between the curves are getting closer. If $\alpha$ and $\beta$ close to the limit 1 or 0, a dramatic change has been observed in density profile.

## I. INTRODUCTION

The automobile is a means of transportation, indispensable in daily life. However, in developing countries, e.g. China, India, Bangladesh and Indonesia [1], m-vehicles come in increasing numbers, and simultaneously no motorized vehicles (thereafter nm-vehicle, including bicycle, three-wheeler, motorcycle) are still prevalent for most short-distance trips due to low income levels or convenient parking.

Modeling road traffic behavior using cellular automata (CA) has become a well-established method to model, analyze, understand, and even forecast the behavior of real road traffic, because the automata's evolution rules are simple, straightforward to understand, computationally efficient and sufficient to emulate much of the behavior of observed traffic flow [2]. The asymmetric simple exclusion process is one of these models. It has acquired paradigmatic status for several reasons: first, with open boundaries, it shows highly nontrivial behaviors such as distinct phases, shocks, and long-range correlations in both space and time; second, in its simplest forms, its steady-state properties, as well as selected dynamic quantities, can be found exactly; and finally, the model is closely related to applications of traffic flow [3].

Despite findings from previous studies, one important point has been missing, which is the decision-making process of drivers, which unequivocally affects traffic flows. Owing to this background, Yamauchi et al. [4] developed a framework integrating a traffic flow model with game theory. And Nakata et al. [5] observe a dilemma structure in the metastable phase by using the stochastic Nishinari–Fukui–Schadschneider (S-NFS) [6] model instead of SOV. Beyond those backgrounds, the objectives of this paper are to add a cooperative game theory framework as a rational decision process to the two-lane bicycle traffic model, construct a new CA model.

## II. MODEL

The model is defined in a two-lane lattice of L×2 sites, where L is the length of a lane. The occupation variable is $\tau_{l,i}$ where $\tau_{l,i} \geq 1$ or $\tau_{l,i} = 0$ means that the state of the i th site in lane A or B is occupied or vacant. Following dynamical rules was applied.

---

[a] chiyu_wang@yahoo.cn
[b] lhwang@gxnu.edu.cn
[c] crh@gxnu.edu.cn

1. Sub case i=1: If $\tau_{l,i} = 0$, a cyclist enters the system with rate $\alpha$ .
2. Sub case i=N. If $\tau_{l,i} \geq 1$, the cyclist leaves the system with rate $\beta$ .
3. Sub case 1<i<N. If $\tau_{l,i} \geq 1$, the cyclist moves into site $(l, i+1)$ with unit rate if $\tau_{l,i+1} = 0$. Otherwise, it changes to the other lane with rate $\omega_1$ if the corresponding site on the other lane is empty.

We carry out the simulations on the lane of length L=400. The first 10 000 steps are discarded to let the transient time die out. The following 2 000 steps are used for averaging.

Note that the model is different from the model of Jiang et al [7] and the model of Pronina and Kolomeisky[8], in which the particles change lane but not overlap. This difference leads to different results.

## III. Results

In this section, we study the situation arising from preference for the right line, in which the lane changing rates $\omega_{left}$ and $\omega_{right}$ are inversely proportional to L. We denote $\omega_{\text{left}} \text{L} = \Omega_{left}$ and $\omega_{\text{right}} L = \Omega_{right}$ . Here $\Omega_{\text{left}}$ and $\Omega_{\text{right}}$ are constants.

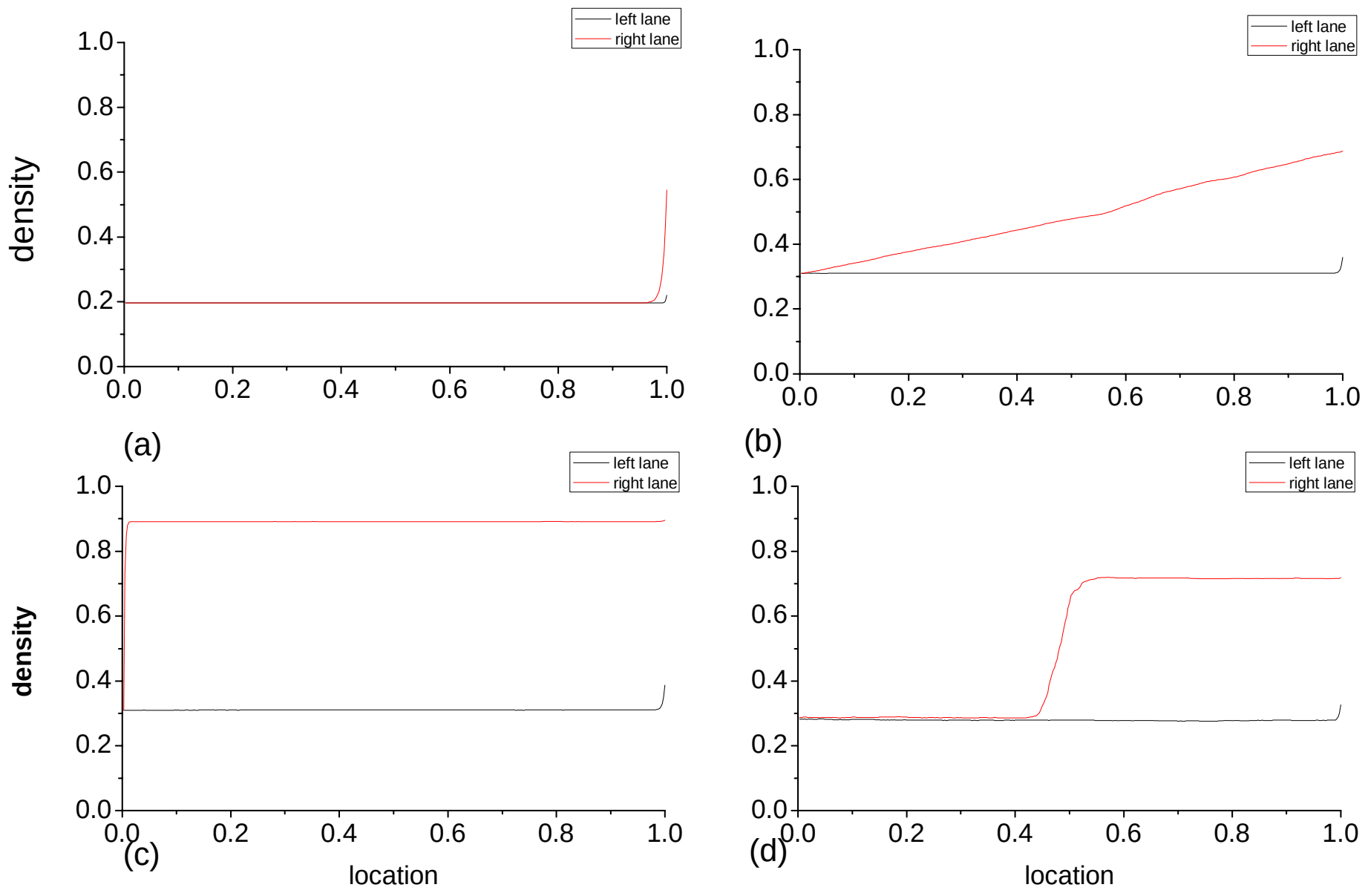

FIG. 1. Typical density profiles. The parameters are $\Omega_{\text{left}} = 1, \Omega_{\text{right}} = 0$ ,L=400. In (a) $\alpha = 0.2, \ \beta = 0.3$ ; (b) $\alpha = 0.4, \ \beta = 0.4$ ;(c) $\alpha = 0.4, \ \beta = 0.25$ ;(d) $\alpha = 0.4, \ \beta = 0.4$ ;The red lines are right lane results and the black lines are left lane results. And (a), (b), (c) is the 100 times the average results; (d) is one random result.

### A. Phase and density profile

Figure 1 shows the typical density profiles in the different conditions. According to our simulation, there are three kinds of state. The first, like (a), belong to free-flow phase that two lanes have the same density except that a sharp increase in the density of two-lane in the exit, and at this time, the right lane density is greater than the left one. When $\alpha < \beta$ in the condition that parameters are $\Omega_{\text{left}} = 1, \Omega_{\text{right}} = 0$ ,which means cyclists totally prefer the right line rules, there are the free-flow phase. The second, like(b), belong to metastable phase that density of right lane showing the growth with the form of straight line, but the density of the left lane do not change.

The reason of this is the localization of shock in the right lane in one time result, like (d). The shock randomly walks from start of road to the end in the right lane. The shock happened more in the exit than in the entrance. That's why the slope is positive. When $\alpha = \beta$ in the condition that parameters are $\Omega_{\text{left}} = 1, \Omega_{\text{right}} = 0$, there are the metastable phase. The third, like (c) belong to jam phase that the shock happened in the entrance, but the density of the left lane do not change. The density profile suggests that even in the jam phase there is a lane not jam in the bike lane. When $\alpha > \beta$ in the condition that parameters are $\Omega_{\text{left}} = 1, \Omega_{\text{right}} = 0$, there are the jam phase.

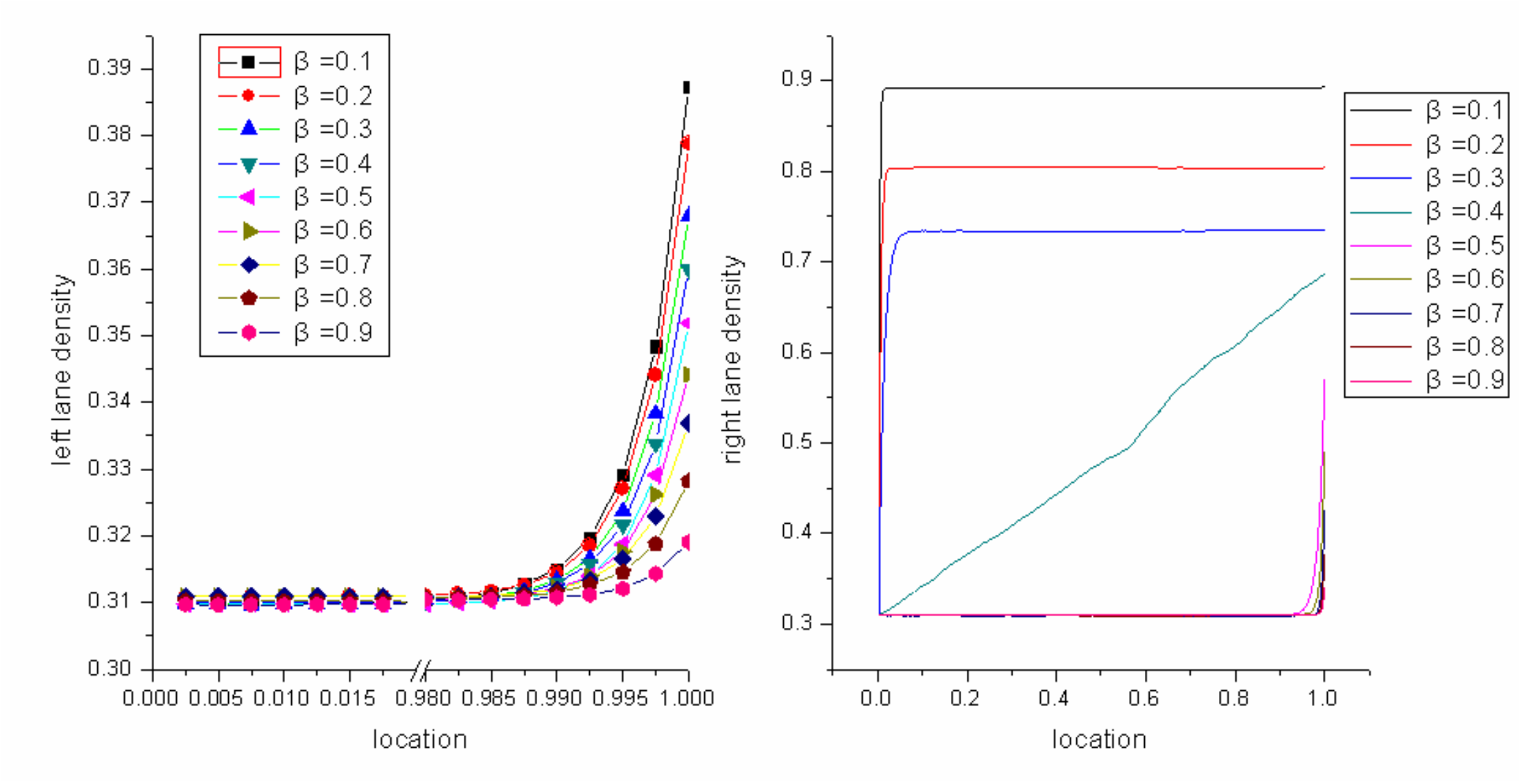

FIG. 2. Average overall density profile in the left lane and right lane for the change of $\beta$, when the $\alpha$ is constant. The parameters are $\Omega_{\text{left}} = 1, \Omega_{\text{right}} = 0$ ,L=400， $\alpha = 0.4$.

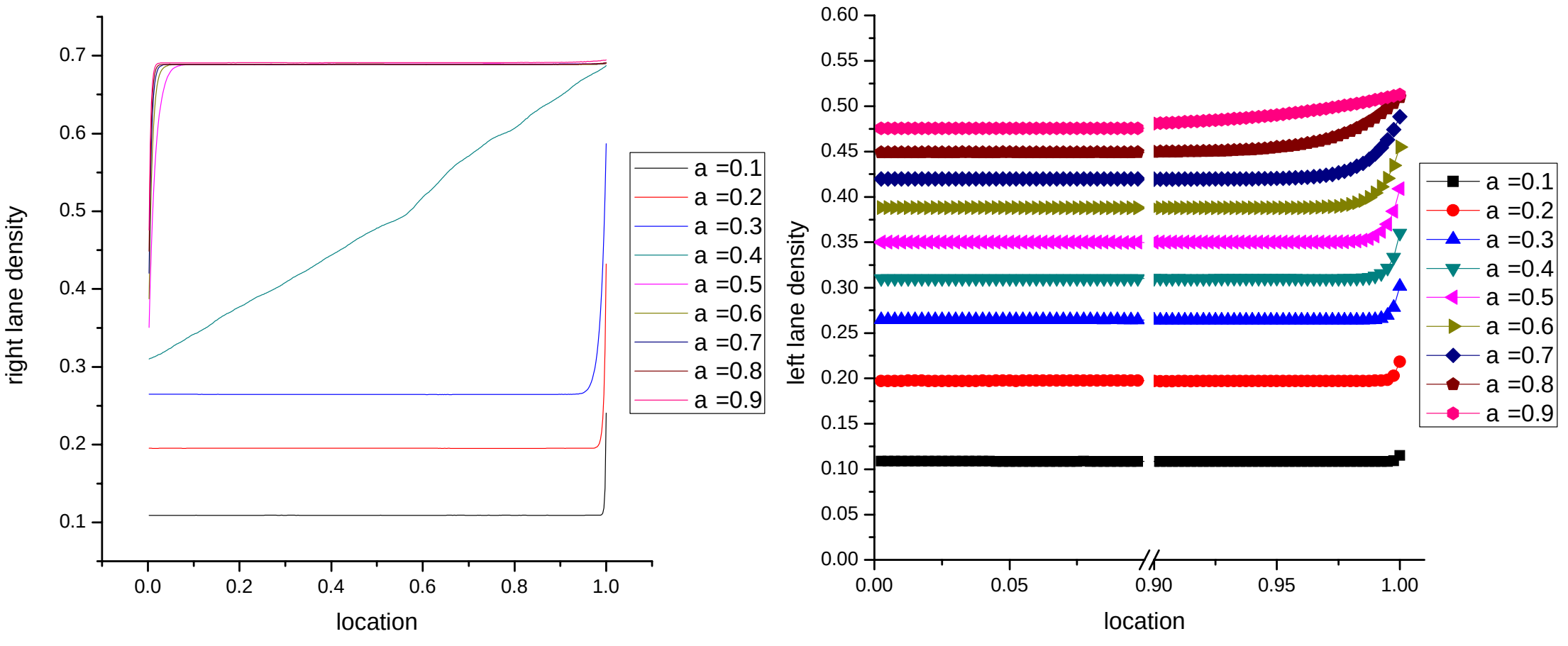

FIG. 3. Average overall density profile in the left lane and right lane for the change of $\alpha$, when the $\beta$ is constant. The parameters are $\Omega_{\text{left}} = 1, \Omega_{\text{right}} = 0$ ,L=400， $\beta = 0.4$.

Figure 2 shows the 100 times average overall density profile in the left lane and right lane for the change of $\beta$, while the $\alpha$ is constant. The left picture indicate that left lanes have almost the same density except that a sharp increase in the density of lane in the exit in the various value of $\beta$. With the increasing of $\beta$, the average overall density in the exit of left lanes are decreasing. On the other word, in the exit of left lanes, the curves never intersect each other and arrange in descending order of $\beta$ with it increasing. The right picture in Figure 2 indicate that right lanes have witnessed phase change from jam phase to metastable phase and then from metastable phase to free-flow phase with the value of $\beta$ increasing. The curves also never intersect each other and arrange in descending order of $\beta$ with it increasing.

Figure 3 shows the 100 times average overall density profile in the left lane and right lane for the change of $\alpha$, while the $\beta$ is constant. The left picture in Figure 3 indicate that right lanes have witnessed phase change from free-flow phase to metastable phase and then from metastable phase to jam phase with the value of $\alpha$ increasing. The curves never intersect each other and arrange in ascending order of $\alpha$ with it increasing. The right picture indicate that in left lanes the density is constant except that a slightly increase in the density of lane in the exit in the various value of $\alpha$. On the other word, in the exit of left lanes, the curves never intersect each other and arrange in ascending order of $\alpha$ with it increasing. And the distance between the two curves are getting closer, from bottom to top.

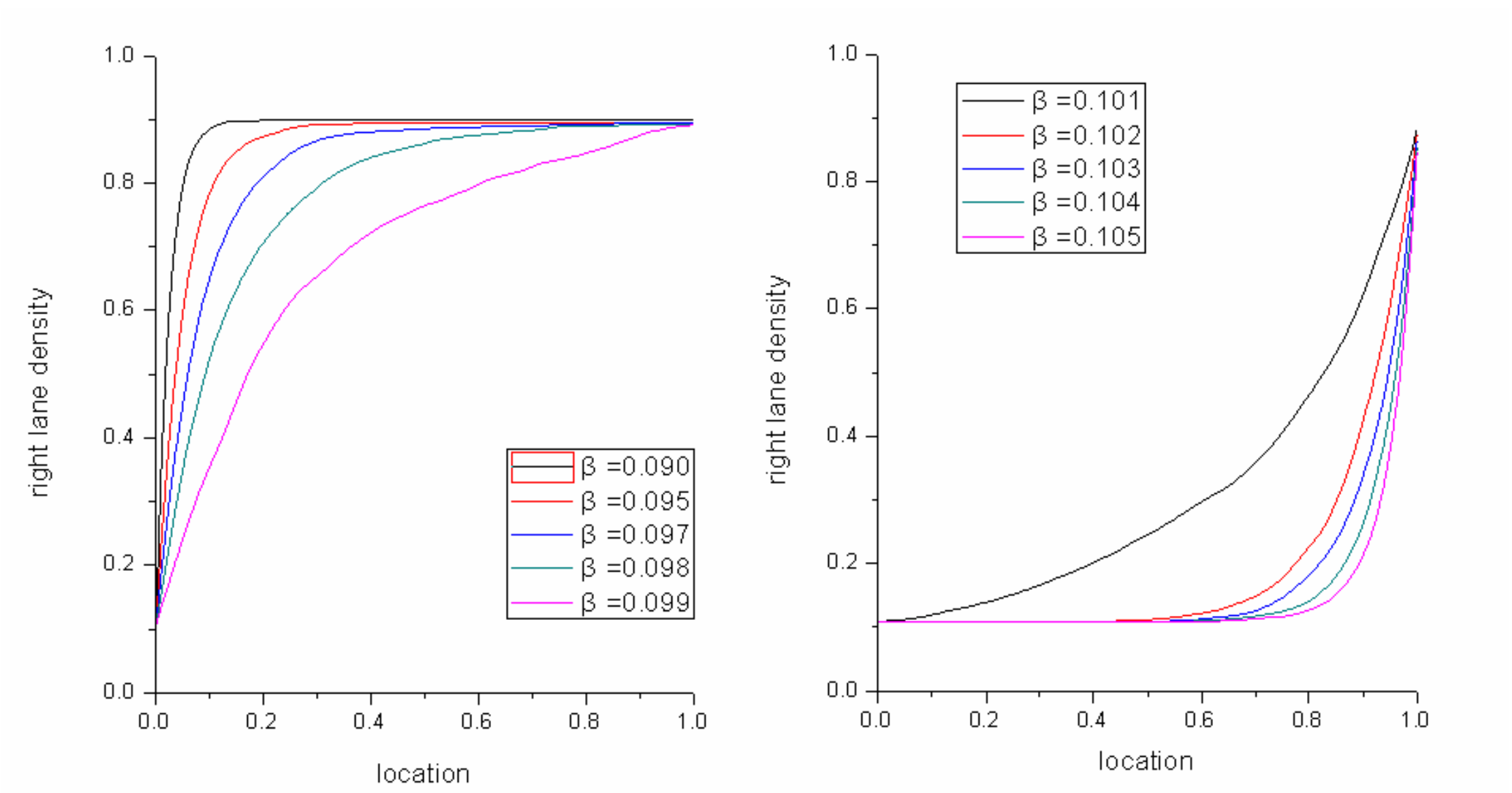

FIG. 4. Average overall density profile in the right lane for the change of $\beta$, when the $\alpha$ is constant. The parameters are $\Omega_{\text{left}} = 1, \Omega_{\text{right}} = 0$ ,L=400, $\alpha = 0.1$.

From Figure 2 and Figure 3, we can see that in the experiment when the $\alpha$ (or $\beta$) is constant and $\beta$ (or $\alpha$) changes in isometric, the distance the two curves show the density changes regularity. When the $\alpha$ is constant and $\beta$ changes in isometric, the distance between the curves are getting closer, from top to bottom in the right lane. When the $\beta$ is constant and $\alpha$ changes in isometric, the distance between the curves are getting closer, from bottom to top in the right lane; in the left lane the distance between the curves are getting closer, from bottom to top ,it just like spectral lines of hydrogen atoms. Figure 4 further study of this phenomenon near the metastable phase ($\Omega_{\text{left}} = 1, \Omega_{\text{right}} = 0$ ,L=400, $\alpha = 0.1$, $\beta = 0.1$) in the right lane. The result is 1000 times average overall density profile. The left picture in Figure 4 indicate that with the value of $\beta$ decreasing, right lanes have witnessed phase change from metastable phase to jam phase and the distance between the curves are getting closer. The right picture in Figure 4 indicate that with the value of $\beta$ increasing, right lanes have witnessed phase change from and the distance between the curves are getting closer. In other words, the distances between the curves are getting closer from metastable phase to the other two phases.

B. Limit state about $\alpha$ and $\beta$

When $\alpha$ and $\beta$ close to the limit 1 or 0, a dramatic change has been observed in density profile. Figure 5 shows that when $\beta$ close to the limit 0 (in there we let $\beta = 0.001$), in right lanes the shock happened not only in the entrance but also in the end, with the value of $\alpha$ increasing, like (b) and (c). At the same time, the value of density in the end is more than 1. When $\alpha$ further come to 1 (in there we let $\alpha = 0.999$), like (d) shows, the curve linear growth with random noise, after the localized shock happened in the beginning.

When $\beta$ close to the limit 1 (in there we let $\beta = 0.999$), most of the results look like random which be showed in Figure 6 (a), (b), (c) and (d), though it is the 1000 times average overall density profile. However, when $\alpha$ further come to 1 (in there we let $\alpha = 0.999$), in the right lane the curve of density linear growth with random noise and then the shock happened in the end of the road which showed in Figure 6 (e); in the right lane the curve of density tardy growth with random pulsation. If we careful observe it with discrete dot, it just likes two curves growth independently. The further study show this phenomenon will happened in the condition that both $\alpha$ and $\beta$ close to the limit 1. The left picture in FIG. 7 show when $\alpha = 1.0$, $\beta = 0.999$, if the curve of density observed with discrete dot, the left lane's curve will become two parallel lines and the right one become a horizontal line and a slash. The right picture in FIG. 7 show when $\alpha = 1.0$, $\beta = 1.0$, if the curve of density observed with discrete dot, the left lane's curve will coincide with the right one.

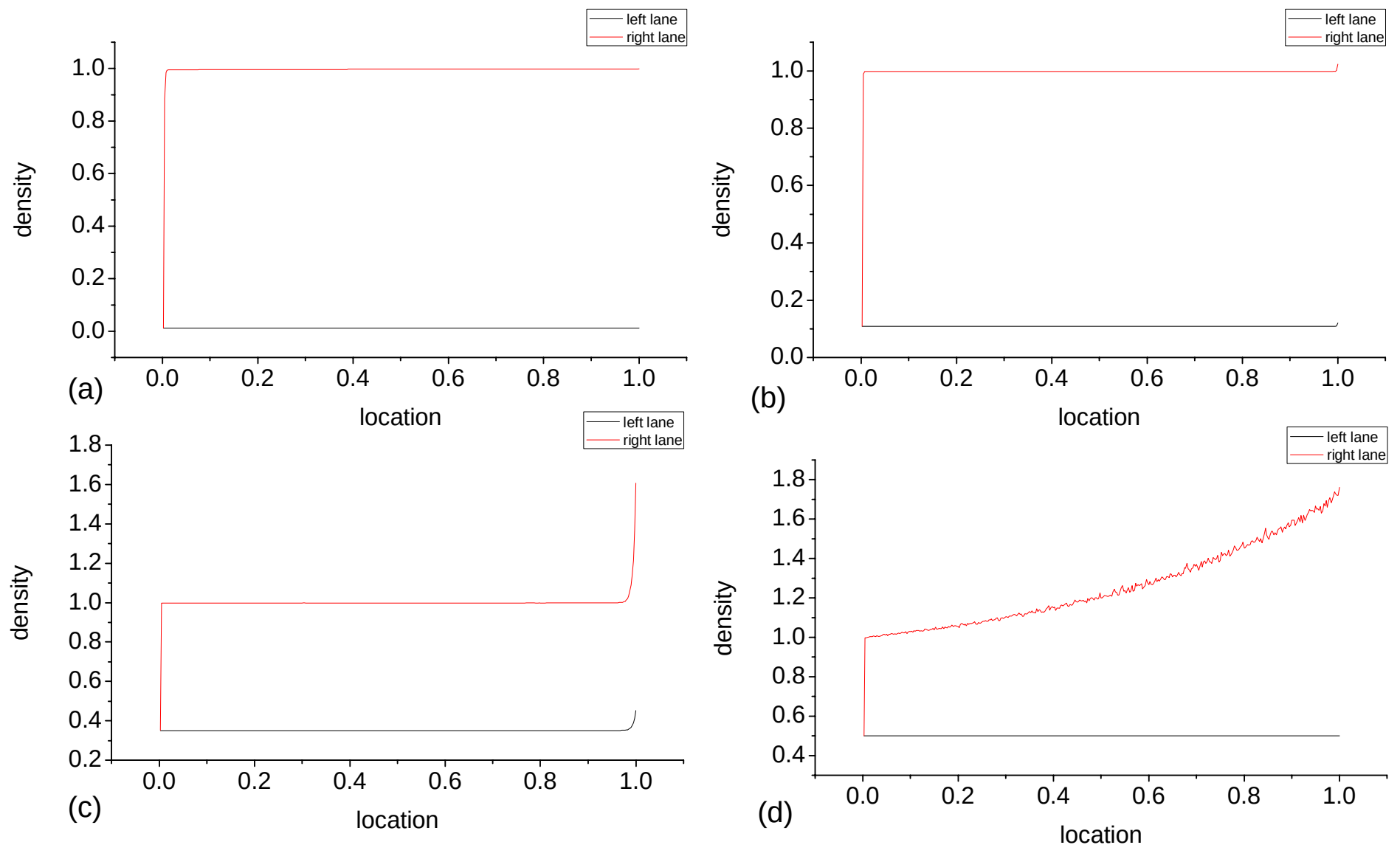

FIG. 5. Average overall density profile. The parameters are $\Omega_{\text{left}} = 1, \Omega_{\text{right}} = 0$ ,L=400, $\beta = 0.001$. In (a) $\alpha = 0.01$; (b) $\alpha = 0.1$;(c) $\alpha = 0.5$;(d) $\alpha = 0.999$. The red lines are right lane results and the black lines are left lane results.

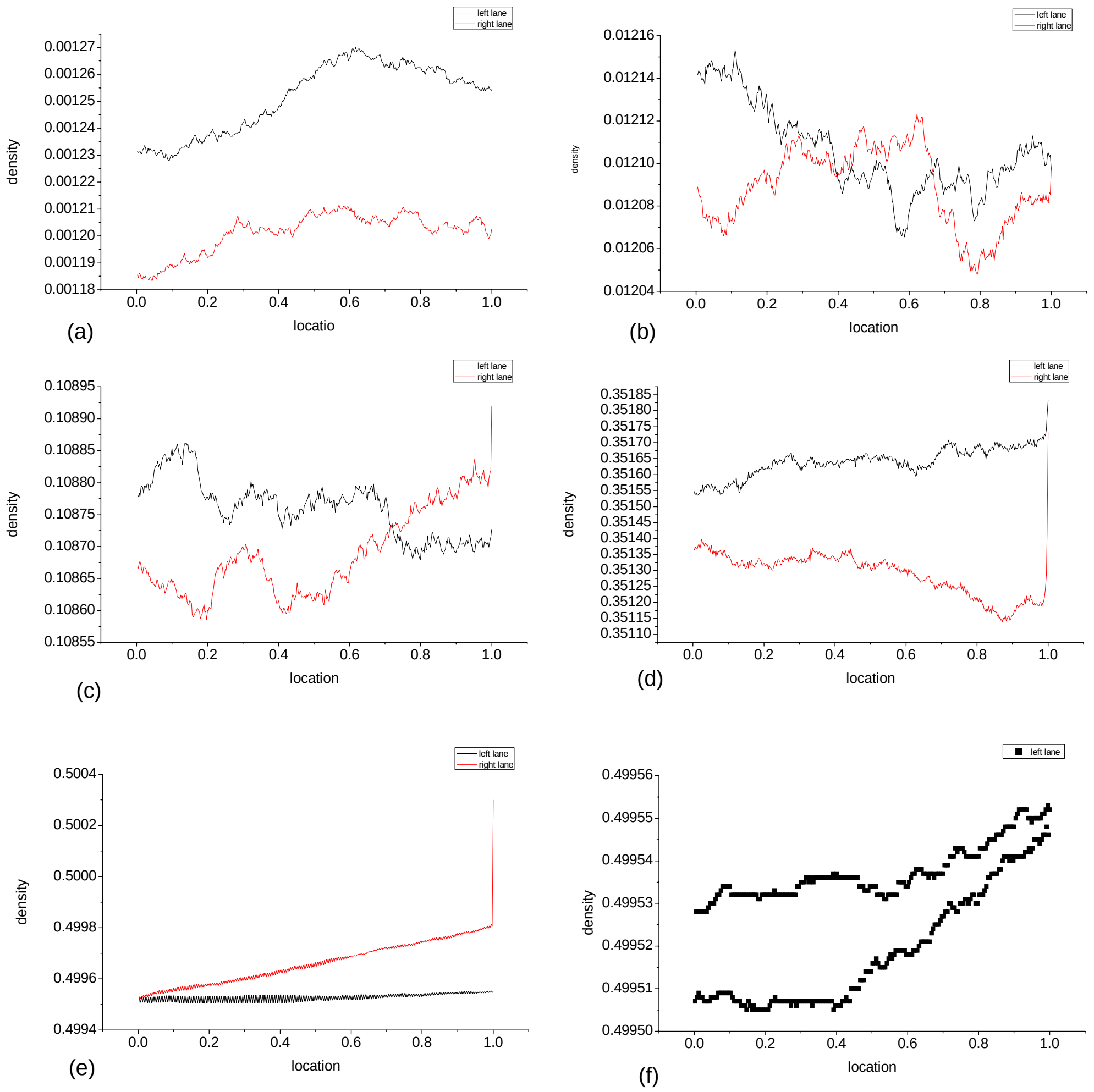

FIG. 6. 1000 times average overall density profile. The parameters are $\Omega_{\text{left}}=1, \Omega_{\text{right}}=0$ ,L=400， $\beta=0.999$ . In (a) $\alpha=0.001$; (b) $\alpha=0.01$; (c) $\alpha=0.1$; (d) $\alpha=0.5$; (e) $\alpha=0.999$; (f) the left lane density result of (e). The red lines are right lane results and the black lines are left lane results.

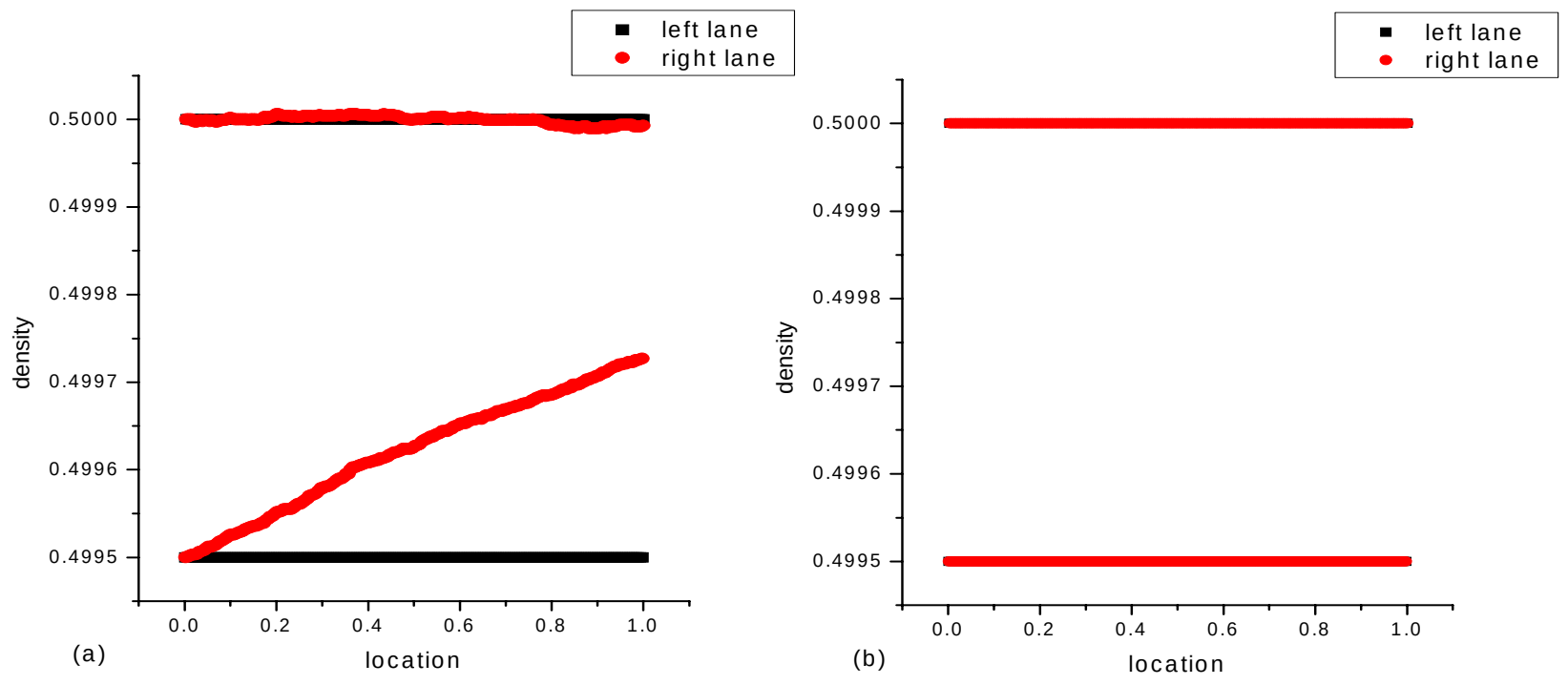

FIG. 7. 1000 times average overall density profile. The parameters are $\Omega_{\text{left}}=1, \Omega_{\text{right}}=0$ ,L=400， $\alpha=1.0$ . In (a) $\beta=0.999$ ; (b) $\beta=1.0$ ; The red lines are right lane results and the black lines are left lane results.

IV. Conclusion

In this paper, we have studied a two-lane totally asymmetric simple exclusion process, in which particles could jump between the two lanes with asymmetric rates. We have investigated the situation about that cyclists totally prefer the right line rules. The density in a certain area appeared surge We have studied the phase boundary, the result show when the $\alpha$ (or $\beta$) is constant and $\beta$ (or $\alpha$) changes in isometric, the distance between the curves are getting closer. If $\alpha$ and $\beta$ close to the limit 1 or 0, a dramatic change has been observed in density profile.

What we present here implies that the cooperative game theorists use may underlie a traffic flow phenomenon that is believed to be a typical physics problem.